\documentclass[12pt,a4paper]{article}%
\usepackage{epsfig}
\usepackage{amsmath}
\usepackage{amsfonts}
\usepackage{amssymb}
\usepackage{graphicx}
\usepackage[T1]{fontenc}
\usepackage[utf8]{inputenc}
\usepackage{cite}
\usepackage{authblk}%
\providecommand{\U}[1]{\protect\rule{.1in}{.1in}}
\begin{document}

\title{A new symmetry theory for non-Hermitian Hamiltonians}
\author{Mustapha Maamache\thanks{E-mail: maamache@univ-setif.dz} and Nour El Houda Absi\\Laboratoire de Physique Quantique et Syst\`{e}mes Dynamiques,\\Facult\'{e} des Sciences, Universit\'{e} Ferhat Abbas S\'{e}tif 1, S\'{e}tif
19000, Algeria. }
\date{}
\maketitle

\begin{abstract}
The main result of the paper is the introduction of the new symmetry operator,
$\tilde{\eta}=\mathcal{PT}\eta$, which acts on the Hilbert space. This
symmetry $\tilde{\eta}$ called $\eta$ pseudo $\mathcal{PT}$ symmetry explores
the conditions under which a non-Hermitian Hamiltonians can possess real
spectra despite the violation of $\mathcal{PT}$ symmetry, that is the adjoint
of $H$, denoted $H^{\dag}$ \ is expressed as $H^{\dag}=\mathcal{PT}%
H\mathcal{PT}$. The $\eta$ pseudo $\mathcal{PT}$ symmetry condition requires
the Hamiltonian to commute with the $\tilde{\eta}$ operator, leading to real
eigenvalues. We discuss some general implications of our results for the
coupled non hermitian harmonic oscillator.

Dedicated to the memory of \textbf{Djabou Zoulikha} mother of \textbf{Mustapha
Maamache}

\end{abstract}

\section{Introduction}

In the framework of modern quantum theory, it is well established that the
time evolution of a quantum system represented by vectors $\left\vert
\psi\right\rangle $ belonging to a Hilbert space $\mathcal{H}$ , where the
inner product is definite positive\textrm{. }The evolution is governed by the
Hamiltonian operator $H$ through the Schr\"{o}dinger equation. This operator,
which is expressed in terms of position, momentum, and other variables,
provides a description of physical quantities such as energy. The principles
of quantum mechanics are based on fundamental postulates, one of which states
that if an Hamiltonian $H$ is Hermitian, its eigenvalues are real.

However, in 1998, Bender and Boettcher \cite{Bender1} discovered\textsf{\ }the
existence of non-Hermitian Hamiltonians that possess real spectra. The reality
of these spectra can be attributed to the symmetry of space-time reflection,
known as $\mathcal{PT}$-symmetry, where the operator $\mathcal{P}$ represents
parity and the operator $\mathcal{T}$ represents time-reversal
\cite{Bender1,Bender4,Bender8', Bender8"} . This concept of $\mathcal{PT}%
$-symmetry has opened the way for further exploration and has been extended to
the notion of pseudo-Hermiticity \cite{Mostafa1,Mostafa7,Mostafa2,Mostafa3}.

Moreover, quantum mechanics is based on a rigorous and systematic mathematical
framework with the introduction of a more extensive category of non-Hermitian
Hamiltonians, commonly referred to as pseudo-Hermitian Hamiltonians
\cite{sholtz}. These particular Hamiltonians, while not self-adjoint, adhere
to the condition of pseudo-Hermiticity, characterized by the relation
$H^{+}=\eta H\eta^{-1}$, where $\eta$ is a linear Hermitian operator called
the metric operator. Furthermore, there be a connection betwen the Hamiltonian
$H$ and its conjugate $H^{+}$through the relation $H^{+}=\mathcal{PT}%
H\mathcal{PT}$ called a pseudo $\mathcal{PT}$-symmetry
\cite{maamache2,maamache4}. \ 

The objective of this work is to restore the Hamiltonian $H$\ based on the
introduction of the new symmetry operator, $\tilde{\eta}=\mathcal{PT}\eta$
(pseudo-Hermiticity and pseudo $\mathcal{PT}$-symmetry), for that we recall
the concepts of $\mathcal{PT}$-symmetry, pseudo-Hermiticity, and pseudo
$\mathcal{PT}$-symmetry. Then, we will discuss the theory of the $\eta$ pseudo
$\mathcal{PT}$-symmetry and apply it to solve the 2D time-independent non
Hermitian coupled harmonic oscillator.

In the next section, we recall the fundamental principles of $\mathcal{PT}%
$-symmetry and pseudo-Hermiticity, we then proceed to revisit the concept of
pseudo $\mathcal{PT}$-symmetry. In section three, we present the fundamental
principles of $\eta$ pseudo $\mathcal{PT}$-symmetry theory In section four we
consider a coupled oscillators with non-Hermitian interaction.

\section{Brief review on $\mathcal{PT}$-symmetry, pseudo-Hermiticity and
pseudo $\mathcal{PT}$-symmetry}

\subsection{$\mathcal{PT}$-symmetry}

The central idea of $\mathcal{PT}$-symmetric quantum theory is to replace the
condition that the Hamiltonian of a quantum theory be Hermitian with the
weaker condition that it possesses space-time reflection symmetry
($\mathcal{PT}$-symmetry). In order to define the $\eta$ pseudo $\mathcal{PT}%
$-symmetry properly, we need to define the operators $\mathcal{P}$,
$\mathcal{T}$ and $\eta$. The parity operator, $\mathcal{P}$, corresponds to a
reflection of the spatial coordinates. This means that $\mathcal{P}$ is a
linear operator that changes the sign of both the position and $the$ momentum
operators,
\begin{equation}
\mathcal{P}\text{ :}\left\{  x\rightarrow-x\text{ \ \ , \ \ }p\rightarrow
-p\text{ \ \ }\right\}  .
\end{equation}
The time-reversal operator must be antilinear, meaning that $\mathcal{T}$ must
performs complex conjugation leaving space unchanged and changes the sign of
the momentum operator,%

\begin{equation}
\mathcal{T}:\left\{  x\rightarrow x\text{ \ \ , \ \ }p\rightarrow-p\text{
\ \ , \ \ }i\rightarrow-i\right\}  .
\end{equation}
Applying any of them twice should leave the system unchanged, and since they
reflect different coordinates, the action of one should not affect the other.
This implies that $\mathcal{P}^{2}$ $=$ $\mathcal{T}$ $^{2}$ $=1$ and that
$\left[  \mathcal{P},\mathcal{T}\right]  =0$, which together imply that
$(\mathcal{PT})^{2}=1$.

\bigskip A Hamiltonian $H$ is then said to be $\mathcal{PT}$ -symmetric if
\begin{equation}
\left[  H,\mathcal{PT}\right]  =0. \label{com}%
\end{equation}
In conventional Hermitian quantum mechanics, the energy eigenvalues are real,
the time evolution is unitary, and if a linear operator $O$ commutes with the
Hamiltonian , then $H$ and $O$ have simultaneous eigenfunctions. However this
may not be true for anti-linear $\mathcal{PT}$ operator. However, since 1998
\cite{Bender1} it has been known that the condition of Hermiticity is not
required for real energy eigenvalues and unitary time evolution. In fact, it
is possible for the Hamiltonian to be non-Hermitian and still possess these
important qualities if there exists an anti-linear symmetry $\mathcal{PT}$
which leaves the Hamiltonian invariant.

\subsubsection{ The $\mathcal{PT}$ inner product}

\ In non-Hermitian case, if (\ref{com}) holds the energy eigenvalues are real
if and only if $\mathcal{PT}$-symmetry is unbroken,\textrm{ }that is, if $H$
and $\mathcal{PT}$ \ have the same eigenvectors $\left\vert \psi
_{n}\right\rangle $. However, it has been realized that such theories can lead
to a consistent quantum theory in a modified Hilbert space. Then, a quantum
theory have been developed in a modified Hilbert space for unbroken
$\mathcal{PT}$-symmetric non-Hermitian systems. In this context It is natural
to consider a $\mathcal{PT}$-inner product for such theories, as the
requirement of Hermiticity is relaxed to $\mathcal{PT}$-symmetry. This
$\mathcal{PT}$-inner product for eigenvectors $\left\vert \psi_{n}%
\right\rangle $ is defined as:%

\begin{equation}
\left\langle \psi_{n},\psi_{m}\right\rangle _{\mathcal{PT}}=\int dx\left[
\mathcal{PT}\text{ }\psi_{n}(x)\right]  \psi_{m}(x)=(-1)^{n}\delta_{nm},
\label{ps2}%
\end{equation}
even with this $\mathcal{PT}$-inner product, the norms of the eigenfunctions
may not be positive definite.

\subsubsection{$\mathcal{CPT}$-symmetry and inner product}

To overcome the issue of negative norm, Bender et al \cite{Bender4} introduced
the linear operator $\mathcal{C}$, with eigenvalues $\pm1$: $\mathcal{C}%
{{}^2}%
=1$. This operator commutes with the $\mathcal{PT\ }$operator but not with the
$\mathcal{P}$\ and $\mathcal{T}$\ operators separately.%

\begin{equation}
\lbrack\mathcal{C},\mathcal{PT}]=0,\ \ \ \ [\mathcal{C},\mathcal{P}%
]\neq0,\ \ \ [\mathcal{C},\mathcal{T}]\neq0, \label{c1}%
\end{equation}
they define a new structure of inner product, known as the $\mathcal{CPT\ }%
$\ inner product%

\begin{equation}
\left\langle f,g\right\rangle _{\mathcal{CPT}}\emph{=}\int\emph{dx}\left[
\mathcal{CPT}\text{ }f(x)\right]  \emph{g(x).} \label{ps3}%
\end{equation}

Like the $\mathcal{PT}$\ inner product (\ref{ps2}), this inner product is also
phase independent and conserved in time. The inner product (\ref{ps3}) is
positive definite because (the operator) $\mathcal{C}$\ contributes $-1$\ when
it acts on states with negative $\mathcal{PT}$\ norm. This new operator
$\mathcal{C}$\ resembles the charge-conjugation operator in quantum field
theory. However, the precise meaning of $\mathcal{C\ }$is that it represents
the measurement of the sign of the $\mathcal{PT}$\ norm in (\ref{ps2}) of an
eigenstate. Specifically\ %

\begin{equation}
\mathcal{C}\psi_{n}(x)=(-1)^{n}\psi_{n}(x). \label{ps4}%
\end{equation}

Indeed, the issue of negative norm is effectively resolved.

\subsection{Pseudo-Hermiticity}

Another possibility to explain the reality of the spectrum is making use of
the pseudo/quasi-Hermiticity transformations, which do not alter the
eigenvalue spectra\textrm{. }It \ was shown by
\cite{Mostafa1,Mostafa7,Mostafa2,Mostafa3} that $\mathcal{PT}$-symmetric
Hamiltonians are only a\ specific class of the general families
of{\normalsize \ }the pseudo-Hermitian operators. A Hamiltonian is said to be
$\eta$ pseudo Hermitian if:%
\begin{equation}
H^{\dagger}=\eta H\eta^{-1}, \label{1'}%
\end{equation}
where $\eta$ is a metric operator. The eigenvalues of pseudo-Hermitian
Hamiltonians are either real or appear in complex conjugate pairs while the
eigenfunctions satisfy bi-orthonormality relations in the conventional Hilbert
space. Due to this reason, such Hamiltonians do not possess a complete set of
orthogonal eigenfunctions and hence the probabilistic interpretation and
unitarity of time evolution \ have not been \ satisfied by these
pseudo-Hermitian Hamiltonians.

However, like the case of $\mathcal{PT}$-symmetric non-Hermitian systems, the
presence of the additional operator $\eta$\ in the pseudo-Hermitian theories
allows us to define a new inner product in the following manner
\begin{equation}
\left\langle \phi\right\vert \left.  \psi\right\rangle _{\eta}=\left\langle
\eta\phi\right\vert \left.  \psi\right\rangle =\int\left(  \eta\phi(x)\right)
\psi(x)dx=\left\langle \phi\right\vert \left.  \eta\psi\right\rangle .
\label{IP}%
\end{equation}
It is worth noting that the metric operator $\eta$ is not unique, and for each
Hamiltonian $H$\ , there exists an infinite set of such operators. The
specific choice of the pseudo-metric operator $\eta$ determines the
pseudo-Hermiticity of $H$\ .

\subsection{ Pseudo parity-time (pseudo-$\mathcal{PT}$\ )-symmetry\textbf{\ }}

An alternative way to express the adjoint of $H$, denoted as $H^{\dag}$, is as
follows \cite{maamache2,maamache4}%
\begin{equation}
H^{\dag}\mathcal{=PT}H\mathcal{PT}, \label{H+}%
\end{equation}
where in the expression of the inner product (\ref{IP}), the metric\textbf{\ }%
$\eta$\textbf{\ \ }is replaced by $\mathcal{PT}.$ It has been observed that
certain systems do not exhibit exact $\mathcal{PT}$-symmetries, but they can
manifest a distinct form of pseudo $\mathcal{PT}$-symmetry,that extends beyond
the traditional $\mathcal{PT}$-symmetry. \thinspace These systems,
characterized by non-self-adjoint Hamiltonians $H^{\dag}$, deviate from the
standard Hermitian framework and give rise to complex eigenvalues and
non-unitary time evolution, Similar to $\mathcal{PT}$-symmetry, pseudo
$\mathcal{PT}$-symmetry is characterized by the Eq.(\ref{H+}).

The concept of pseudo $\mathcal{PT}$-symmetry has found wide application in
various domains, including periodically high-frequency driven
systems\cite{xia1}, time periodic non-hermitian Hamiltonian systems
\cite{maamache3}, optical systems \cite{xia2}, and even the Dirac equation
\cite{maamache4}.

In light of the above discussion, an important question motivates our work
here: how might one investigate the possibility of introducing a new symmetry
that commutes with the Hamiltonian in order to restore unbroken symmetry
similar to the $\mathcal{PT}$-symmetric case?

In this paper we answer this question from a new perspective by defining a new
symmetry $\tilde{\eta}$ as a product of the operators $\mathcal{PT}$ and
$\eta$ namely :%
\begin{equation}
\tilde{\eta}=\mathcal{PT}\eta\text{\ }.\text{ }%
\end{equation}

\section{ Theory of the $\eta$ pseudo $\mathcal{PT}$-symmetry}

In order to introduce the $\eta$ pseudo $\mathcal{PT}$-symmetry theory, let us
substitute the expression of $H^{+}$ (\ref{1'}) into (\ref{H+})
\begin{align}
H  &  =\mathcal{PT}\eta H\eta^{-1}\mathcal{PT},\nonumber\\
&  =\tilde{\eta}H\tilde{\eta}^{-1}, \label{NN}%
\end{align}
this allows us to define a symmetry operator $\tilde{\eta}$ such that%

\begin{equation}
\tilde{\eta}=\mathcal{PT}\eta\text{\ },\text{ }%
\end{equation}
the result (\ref{NN}) indicated the possibility to compensate the broken
symmetry of a Hamiltonian by the presence of $\tilde{\eta}$-symmetry. The
spectra of many non-hermitian Hamiltonian $H$ are indeed real if they are
invariant under the action of metric $\tilde{\eta}$, i.e, $\left[
H,\tilde{\eta}\right]  =0$ and if the energy eigenstates are invariant under
the operator $\tilde{\eta}$.

Therefore, the equation (\ref{NN}) can be reformulated as follows%

\begin{equation}
H=\tilde{\eta}H\tilde{\eta}^{-1}. \label{nor}%
\end{equation}

The operator $\tilde{\eta}$ possesses the following properties :

1) It is invertible \qquad\
\begin{equation}
\tilde{\eta}^{-1}=\eta^{-1}\text{\ }\mathcal{PT}.\text{ }%
\end{equation}

\bigskip2) If we express $H$ in the right side of the equation (\ref{nor}) in
terms of $H^{+}$, we obtain the time-independent quasi-Hermiticity equation%
\begin{align}
\mathcal{PT}\eta^{-1}\mathcal{PT} &  =\eta\nonumber\\
&  =\eta^{-1}H^{+}\eta,
\end{align}

which allows us to identify the action of the operator $\mathcal{PT}$ on the
metric $\eta$ and to relate it to its inverse $\eta^{-1}$%

\begin{equation}
\mathcal{PT}\eta\mathcal{PT}=\eta^{-1}, \label{nor1}%
\end{equation}

and%

\begin{equation}
\mathcal{PT}\eta^{-1}\mathcal{PT}=\eta, \label{nor2}%
\end{equation}

multiplying the equation (\ref{nor1}) by $\mathcal{PT}$ on the right and the
equation (\ref{nor2}) on the left, we find%

\begin{equation}
\tilde{\eta}=\tilde{\eta}^{-1}.
\end{equation}

3) $\tilde{\eta}$ commutes with the Hamiltonian $H$\ \
\begin{equation}
\lbrack H,\tilde{\eta}]=0,
\end{equation}

4) $\tilde{\eta}\neq\tilde{\eta}^{+}$where%

\begin{equation}
\tilde{\eta}^{+}=\eta\mathcal{PT}\text{\ },
\end{equation}

and%

\begin{equation}
\text{\ }\tilde{\eta}%
{{}^2}%
=1.
\end{equation}

5) Now, a "natural inner product"\ of functions $\psi_{n}$\ associated with
$\eta$ pseudo $\mathcal{PT}$-symmetric systems is proposed%

\begin{equation}
\left(  \psi_{n},\psi_{m}\right)  _{\tilde{\eta}}=\int dx\left[  \tilde{\eta
}\psi_{n}(x)\right]  \psi_{m}(x)=(-1)^{n}\delta_{nm}, \label{IP1}%
\end{equation}

this inner product implies that energy eigenstates can have a negative
norm,\textrm{\ }%
\begin{equation}
\left(  \psi_{n},\psi_{m}\right)  _{\tilde{\eta}}=(-1)^{n}\delta_{nm}.
\end{equation}

6) In an attempt to extend quantum mechanics to systems with $\eta$ pseudo
$\mathcal{PT}$-symmetry, a remedy against the indefinite metric in Hilbert
space we propose in the form of a linear charge operator $\mathcal{C}$. Then,
the redefined inner product\textrm{\ }%
\begin{equation}
\left(  \psi_{n},\psi_{m}\right)  _{\mathcal{C}\tilde{\eta}}=\int dx\left[
\mathcal{C}\tilde{\eta}\text{ }\psi_{n}(x)\right]  \psi_{m}(x)=\delta_{mn},
\label{IP2}%
\end{equation}

where
\begin{equation}
\mathcal{C}\psi_{n}(x)=(-1)^{n}\psi_{n}(x)\text{.}%
\end{equation}

We note that $\mathcal{C}$ commutes with $H$ and $\tilde{\eta}$, indeed if we
take $\mathcal{C}=\mathcal{P}\tilde{\eta}$, we can easily demonstrate%

\begin{equation}
\left[  \mathcal{C},\tilde{\eta}\right]  =0.
\end{equation}

7) One possibility to explain the reality of the spectrum is to relate the
non-Hermitian Hamiltonian, $H\neq H^{\dag}$, to a Hermitian Hamiltonian, $h$
$=$ $h^{+}$, through the action of a map $\rho$ such that%

\begin{equation}
H=\mathcal{PT}\rho^{\dag}h\rho^{-1\dag}\mathcal{PT}, \label{hh}%
\end{equation}

where $\eta=\rho^{\dag}\rho$ is the time-independent metric. Thus, we deduce
that the conjugate of the Dyson operator $\rho$ satisfies \ $\rho^{\dag
}=\mathcal{PT}\rho^{-1}\mathcal{PT}.$ The eigenvectors of $h$\textrm{ }denoted
as\textrm{ }$\left\vert \psi_{n}^{h}\right\rangle $, can be chosen in such a
way that they are also the eigenvectors of $\mathcal{PT}$\textrm{ ,
}since\textrm{ }$\left[  h,\mathcal{PT}\right]  =0$.

\section{\bigskip Application: 2D Coupled Harmonic Oscillator}

We examine a system of 2D coupled oscillators described by the Hamiltonian%
\begin{equation}
H(t)=\frac{1}{2}\overset{2}{\underset{i=1}{\sum}}\left[  P_{i}^{2}+c%
{{}^2}%
_{i}X_{i}^{2}\right]  +\frac{1}{2}ic_{3}X_{1}X_{2}, \label{CH}%
\end{equation}
this complex Hamiltonian is pseudo $\mathcal{PT}$-symmetric because $i$
changes sign under time reversal $\mathcal{T}$ and it is assumed that every
coordinate $X_{\substack{1\\\text{ \ }}}$and $X_{2}$ changes sign under parity
$\mathcal{P}$ \cite{asiri}.\ However, the authors of Refs. \cite{bahbani,
beygi, fring, fring1} consider the case where $H$ (\ref{CH}) is partially
$\mathcal{PT}$-symmetric, that is, it remains invariant if the sign of $i$
changes and simultaneously reverse the sign of only the
$X_{\substack{1\\\text{ \ }}}$or $X_{2}$ coordinates.

In order to solve this quantum system ,we introduce the metric operator in the form%

\begin{equation}
\eta=\exp\left[  \theta\left(  P_{2}X_{1}-P_{1}X_{2}\right)  \right]  ,
\end{equation}

its inverse is given by%

\begin{equation}
\eta^{-1}=\exp\left[  -\theta\left(  P_{2}X_{1}-P_{1}X_{2}\right)  \right]  ,
\end{equation}

\bigskip under which the operators $P_{1}$, $P_{2}$, $X_{1}$ and $X_{2}$
transform such as
\begin{equation}
\eta\left(  P_{1}\right)  \eta^{-1}=P_{1}\cosh\theta+iP_{2}\sinh\theta,
\end{equation}%
\begin{equation}
\eta\left(  P_{2}\right)  \eta^{-1}=P_{2}\cosh\theta-iP_{1}\sinh\theta,
\end{equation}%
\begin{equation}
\eta\left(  X_{1}\right)  \eta^{-1}=X_{1}\cosh\theta+iX_{2}\sinh\theta,
\end{equation}%
\begin{equation}
\eta\left(  X_{2}\right)  \eta^{-1}=X_{2}\cosh\theta-iX_{1}\sinh\theta,
\end{equation}

therefore, $\eta H\eta^{-1}$is transformed as%

\begin{equation}
\eta H\eta^{-1}=H^{+}+\frac{1}{2}\left[  \left(  c%
{{}^2}%
_{1}-c%
{{}^2}%
_{2}\right)  \sinh\theta+c_{3}\cosh\theta\right]  \left(  \sinh\theta\left(
X_{1}^{2}-X_{2}^{2}\right)  +\text{ }2iX_{1}X_{2}\cosh\theta\right)  ,
\label{a}%
\end{equation}

\bigskip applying on the both side of equation (\ref{a}) the $\mathcal{PT}$
operator, we obtain%

\begin{equation}
\tilde{\eta}H\tilde{\eta}^{-1}=H+\frac{1}{2}\left[  \left(  c%
{{}^2}%
_{1}-c%
{{}^2}%
_{2}\right)  \sinh\theta+c_{3}\cosh\theta\right]  \left(  \sinh\theta\left(
X_{1}^{2}-X_{2}^{2}\right)  -\text{ }2iX_{1}X_{2}\cosh\theta\right)  ,
\label{a1}%
\end{equation}

For the Hamiltonian $\tilde{\eta}H\tilde{\eta}^{-1}$ (\ref{a1}) to be $H$ we
need to impose that%

\begin{equation}
\left[  \left(  c%
{{}^2}%
_{1}-c%
{{}^2}%
_{2}\right)  \sinh\theta+c_{3}\cosh\theta\right]  =0,
\end{equation}

therefore%

\begin{equation}
\theta=\tanh^{-1}\left[  \frac{c_{3}}{\left(  c_{2}^{2}-c_{1}^{2}\right)
}\right]  .
\end{equation}

Using standard techniques from $\mathcal{PT}$ -symmetric/quasi-Hermitian
quantum mechanics , it can be decoupled easily into two harmonic oscillators
\cite{asiri, bahbani, beygi, fring, fring1} by applying a Dyson map operator
$\eta^{1/2}=\exp\left[  \frac{\theta}{2}\left(  P_{2}X_{1}-P_{1}X_{2}\right)
\right]  $ on $H$ , \textbf{that is}%
\begin{equation}
h=\eta^{1/2}H\eta^{-1/2}=\frac{1}{2}\left(  P_{1}^{2}+P_{2}^{2}\right)
+\frac{1}{2}\left(  \omega_{1}^{2}X_{1}^{2}+\omega_{2}^{2}X_{2}^{2}\right)
\label{2.1}%
\end{equation}
\textbf{where} %

\begin{equation}
\omega_{1}^{2}=\frac{c_{1}\cosh%
{{}^2}%
(\frac{\theta}{2})+c_{2}\sinh%
{{}^2}%
(\frac{\theta}{2})}{\cosh(\theta)}\text{ \ \ , \ }\omega_{2}^{2}=\frac
{c_{2}\cosh%
{{}^2}%
(\frac{\theta}{2})+c_{1}\sinh%
{{}^2}%
(\frac{\theta}{2})}{\cosh(\theta)}\text{ , }\label{2.2}%
\end{equation}
\textbf{by this last observation it follows that one has a real energy
eigenvalues for the Hamiltonian} $H$ (\ref{CH}) (or (\ref{a1}))%
\begin{equation}
E_{_{n}}=\omega_{1}\left(  n_{1}+\frac{1}{2}\right)  +\omega_{2}\left(
n_{2}+\frac{1}{2}\right)  .
\end{equation}
\textbf{provided that} $\left(  c_{2}^{2}-c_{1}^{2}\right)  ^{2}>c_{3}^{2}$,
\textbf{and its corresponding eigenfunctions are}

\begin{equation}
\left\vert \psi_{n}\right\rangle =\mathcal{PT}\eta^{1/2}\left\vert \phi
_{n}^{os}\right\rangle \text{ }%
\end{equation}
\textbf{where}
\begin{equation}
\phi_{n}^{os}(x)=\frac{\left(  \omega_{1}\omega_{2}\right)  ^{\frac{1}{4}}%
}{\sqrt{\pi\text{ }2^{n_{1}+n_{2}}n_{1}!n_{_{2}}!}}e^{\left[  -\left(
\frac{\omega_{1}}{2}X_{1}%
{{}^2}%
+\frac{\omega_{2}}{2}X_{2}%
{{}^2}%
\right)  \right]  }H_{n_{1}}\left[  \omega_{1}^{1/2}X_{1}\right]  H_{n_{2}%
}\left[  \omega_{2}^{1/2}X_{2}\right]  ,
\end{equation}
\textbf{expressed in function of Hermite polynomial} $H_{n_{i}}\left[
\omega_{i}X_{i}\right]  ,$ \textbf{are eigenfunctions of the separable
Hamiltonian. Thus the eigenfunctions} $\psi_{n\text{ }}$\textbf{of the
Hamiltonian} $H$ are%
\begin{equation}
\psi_{n}(x)=\frac{\left(  \omega_{1}\omega_{2}\right)  ^{\frac{1}{4}}}%
{\sqrt{\pi\text{ }2^{n_{1}+n_{2}}n_{1}!n_{_{2}}!}}\mathcal{PT}\left\{
\eta^{1/2}e^{\left[  -\left(  \frac{\omega_{1}}{2}X_{1}%
{{}^2}%
+\frac{\omega_{2}}{2}X_{2}%
{{}^2}%
\right)  \right]  }H_{n_{1}}\left[  \omega_{1}^{1/2}X_{1}\right]  H_{n_{2}%
}\left[  \omega_{2}^{1/2}X_{2}\right]  \right\}  .\text{ }%
\end{equation}
\textbf{Having introduced the operator} $\mathcal{C}$\textbf{ in the last
subsection , we can now use the new} $\mathcal{C}\tilde{\eta}$ \textbf{inner
product defined in} (\ref{IP2})%

\begin{align}
\left\langle \psi_{n}\right.  \left\vert \psi_{m}\right\rangle _{\mathcal{C}%
\tilde{\eta}} &  =\int\left[  \mathcal{C}\tilde{\eta}\psi_{n}\right]  ^{\ast
}\psi_{m}dX_{1}dX_{2},\nonumber\\
&  =\int\left[  \eta^{1/2}\phi_{n}^{os}\right]  ^{\ast}\eta^{-1/2}\phi
_{n}^{os}dX_{1}dX_{2},
\end{align}
where%

\begin{equation}
\mathcal{C}\tilde{\eta}\psi_{n}(x)=(-1)^{n}\tilde{\eta}\psi_{n}(x).
\end{equation}

\textbf{By puting the change of variables} $U=\sqrt{\omega_{1}}X_{1}$
\textbf{and} $V=\sqrt{\omega_{2}}X_{2}$ , \textbf{we find that the
eigenfunctions} $\psi_{n}(x)$ \textbf{are orthogonal with respect to this
scalar product}%

\begin{equation}
\left\langle \psi_{n}\right.  \left\vert \psi_{m}\right\rangle _{\mathcal{C}%
\tilde{\eta}}=\delta_{nm}.
\end{equation}

\section{Conclusion}

In the context of non-Hermitian time-independent quantum mechanics many
systems are known to possess real spectra. This observation leads to the
concept of space-time reflection symmetry, known as $\mathcal{PT}$-symmetry.

We have demonstrated that by introducing metric operator $\eta$ into a pseudo
$\mathcal{PT}$ non-Hermitian Hamiltonian, that is $H^{\dag}\mathcal{=PT}%
H\mathcal{PT}$, we can make the pseudo $\mathcal{PT}$-symmetric regime
physically significant. This allows us to define a new symmetric operateur
$\tilde{\eta}$ that commutes with the Hamiltonian, providing our system with
the same properties as in $\mathcal{PT}$-symmetric systems cases.$.$ We have
shown how to construct quantum theories based on $\tilde{\eta}$ symmetric
non-Hermitian Hamiltonians. In contrast to the Hermitian case, the inner
product for a quantum theory defined by Eq.(\ref{IP1})\ can have a negative
norm. To overcome the issue of negative norm, we define a new structure of
inner product Eq.(\ref{IP2}), called as the $\mathcal{C}\tilde{\eta}$ \ inner
product. We end by solving the 2D coupled non-Hermitian harmonic oscillator.

Finally, we draw the reader's attention to the papers. \cite{Ba, Sol} in which
it was emphasized that the twin concepts of $\eta$-pseudo-hermiticity and weak
pseudo-hermiticity are complementary.

\paragraph{Acknowledgements}

We thank Professor Bijan Bagchi for bringing his article to our attention and
providing us with his comments. We would like to thank Dr. N.\ Amaouche, Dr.
R.\ Zerimeche and H.\ Lahreche for their valuable discussions.

\end{document}